\documentclass[]{spie}  
\usepackage[colorlinks=true]{hyperref}
\usepackage[]{graphicx}
\usepackage{textcomp}
\usepackage{mathptmx}
\usepackage[version=4]{mhchem}
\usepackage{amsmath,amssymb,textcomp,graphicx,url}

\title{Antireflective nanotextures for\\ monolithic perovskite-silicon tandem solar cells} 

\author{Klaus J\"{a}ger,\supit{a, b} Phillip Manley,\supit{a, b} Duote Chen,\supit{a, b} Philipp Tockhorn,\supit{a} David Eisenhauer,\supit{a}\\ Grit K\"{o}ppel,\supit{a} Martin Hammerschmidt,\supit{b, c} Sven Burger,\supit{b, c} Steve Albrecht,\supit{a} and Christiane Becker\supit{a}
\skiplinehalf
\supit{a}Helmholtz-Zentrum Berlin f\"{u}r Materialien und Energie GmbH, Kekul\'{e}str.\ 5, D-12489 Berlin\\
\supit{b}Zuse Institute Berlin, Takustra\ss e 7, D-14195 Berlin\\
\supit{c}JCMwave GmbH, Bolivarallee 22, D-14050 Berlin}

\authorinfo{Send correspondence to K.J.: E-mail: \href{mailto:klaus.jaeger@helmholtz-berlin.de}{klaus.jaeger@helmholtz-berlin.de}}

\begin{document} 
\maketitle 

\noindent This  paper  is  published  in  \emph{Proc.\  SPIE} \textbf{10688}, p.\ 10688A, May 2018 (Photonics for Solar Energy Systems VII, edited by Ralf B.\ Wehrspohn and Alexander N.\ Sprafke, doi: \href{https://doi.org/10.1117/12.2306142}{10.1117/12.2306142})  and  is  made  available  as  an  electronic  preprint  with  permission  of  SPIE.  One  print  or electronic copy may be made for personal use only.  Systematic or multiple reproduction, distribution to multiple locations  via  electronic  or  other  means,  duplication  of  any  material  in  this  paper  for  a  fee  or  for  commercial purposes,  or modification of the content of the paper are prohibited.

\begin{abstract}
Recently, we studied the effect of hexagonal sinusoidal textures on the reflective properties of perovskite-silicon tandem solar cells using the finite element method (FEM). We saw that such nanotextures, applied to the perovskite top cell, can strongly increase the current density utilization from 91\% for the optimized planar reference to 98\% for the best nanotextured device (period 500 nm and peak-to-valley height 500~nm), where 100\% refers to the Tiedje-Yablonovitch limit.\footnote{D.~Chen {\em et~al.}, {\em J.\ Photonics Energy}~{\bf 8}, p.~022601, 2018.}

In this manuscript we elaborate on some numerical details of that work: we validate an assumption based on the Tiedje-Yablonovitch limit, we present a convergence study for simulations with the finite-element method, and we compare different configurations for sinusoidal nanotextures.
\end{abstract}



\section{INTRODUCTION}
\label{sec:intro}  

The power conversion efficiency (PCE) of single-junction silicon solar cells, which are by far the most dominant photovoltaic technology, is theoretically limited to 29.4\%\cite{richter:2013}. Currently, perovskite-silicon tandem solar cells are the most investigated concept to overcome this limit, because they allow for the reduction of thermalization losses at short wavelengths. Because of their steep absorption edge, tuneable bandgap, and high demonstrated solar cell efficiencies, perovskite materials are an ideal candidate as a top-cell material\cite{lal:2014}.

For monolithic tandem solar cells, the power output is maximal when the current densities generated by the top and bottom cells are matched, hence if the available light is distributed equally between the two cells. For perovskite-silicon tandem cells with a wafer-based silicon bottom cell, current matching is mainly achieved by adjusting the thickness of the perovskite layer. For a planar device design, a global optimization of the layer thicknesses in the top cell can be easily performed. Such an optimization not only allows current matching to be reached, but also for the reflective losses of the solar cell to be minimized via optimizing the mutual interference of the thin-film layer stack. However, even after this optimization about 7 mA/cm$^2$ are lost by reflection, which is about 15\% of the totally available current density in the 350--1200 nm wavelength range\cite{jaeger:2017pero}.

In a recent study we investigated the effect of nanotextures in the perovskite layer stack on the reflective properties of the tandem solar cells\cite{chen:2018}. We used hexagonal sinusoidal nanotextures that have proven to strongly reduce reflection at silicon-oxide silicon interfaces \cite{jaeger:2016opex, koeppel:2016, koeppel:2017, jaeger:2018opex}. 

 We performed the optical simulations with the finite element method (FEM). Newton`s method allowed us to obtain the optimal perovskite thickness with very few iterations. We investigated three configurations: (1) all interfaces on top of the silicon wafer are textured; (2) all interfaces between the silicon wafer and the perovskite layer are textured; and (3) all interfaces on top of the perovskite layer are textured. The fully textured device showed the strongest antireflective effect. We chose a nanotexture period of 500 nm, for which the reflective losses can be reduced by more than 2 mA/cm$^2$ with respect to the planar reference in a current-matched tandem device, leading to a maximal achievable current density exceeding 20.2 mA/cm², which translates to a potential PCE increase of about 1.4\% (absolute). For these results the peak-to-valley height of the nanotexture is 250 nm.

We also demonstrated perovskite layers, which were spin-coated onto a sinusoidally nanotextured substrate. According to these results, configuration (2) currently seems most feasible for experimental realization. For this configuration the maximal achievable current density is still about 19.7 mA/cm$^2$ in a current-matched tandem device, which is 0.5 mA/cm$^2$ higher than for the flat reference, equivalent to a potential PCE increase of 0.7\% (absolute).

In this manuscript we explain some details of the work presented in Ref.\ \citenum{chen:2018} in more depth. After explaining the numerical methods in Section \ref{sec:methods}, we elaborate on these topics in Section \ref{sec:results}: (1) checking the validity of applying the Tiedje-Yablonovitch limit for light trapping in the silicon subcell; (2) a convergence study for the mesh dimensions and used degree of polynomials in the FEM dimension; and (3) comparing different configurations of the hexagonal nanotexture.

\section{METHODS}
\label{sec:methods}

\subsection{Details of the Solar Cell Simulations}
\begin{figure}
\centering
\includegraphics{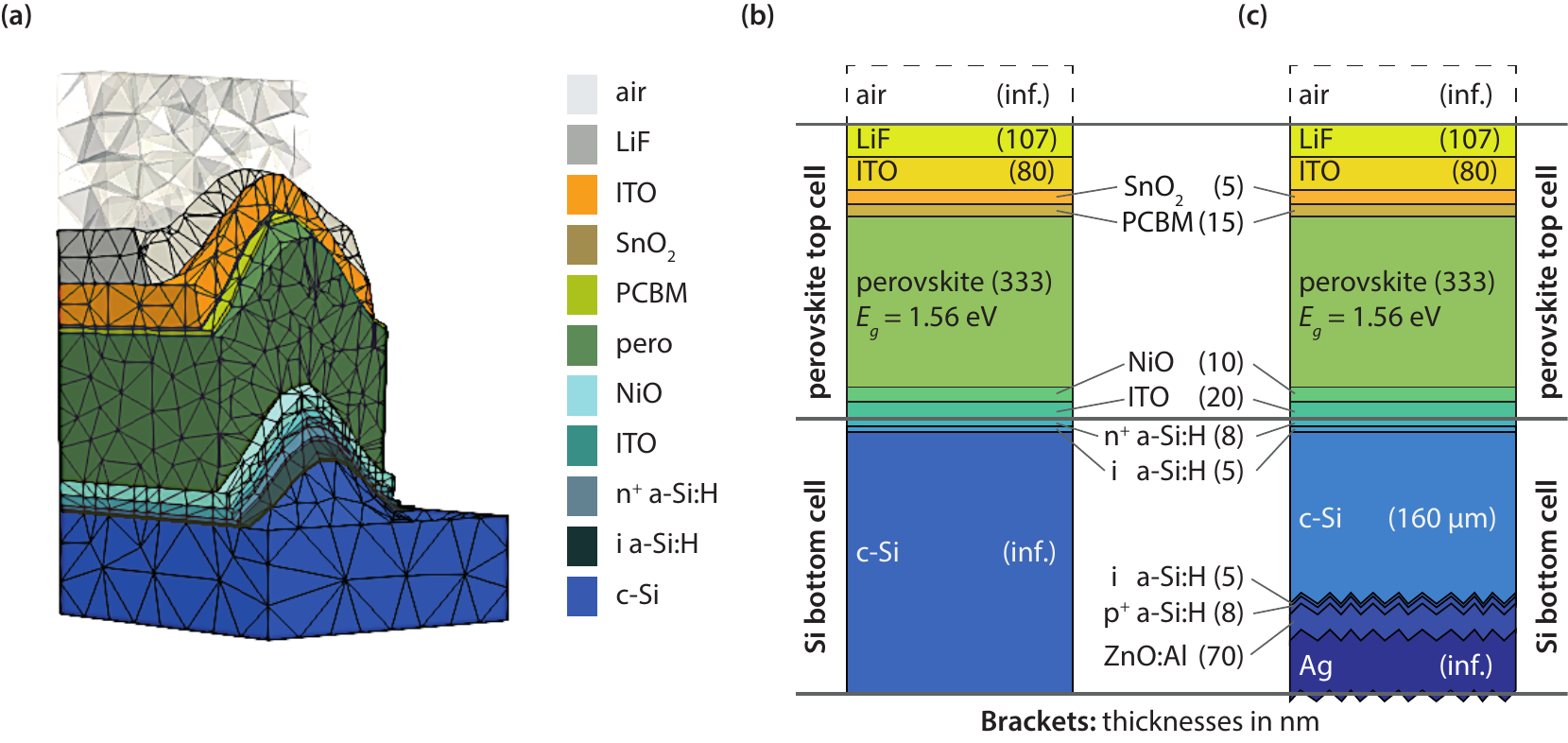}
\caption{(a) The cell architecture considered in the FEM simulations with a double-side textured top cell. This figure shows the \emph{negative cosine} nanotexture, for the \emph{positive cosine} texture the orientation of the texture is reversed. For the FEM simulations, the upper and lower simulation domains are enclosed by air and silicon halfspaces, respectively. The light is incident from the air half space (top). \newline (b) Illustrating the layer thicknesses in our simulated tandem cell. These are the set of optimal layer thicknesses numerically calculated in Ref.\ \citenum{jaeger:2017pero}. The perovskite thickness has been adapted to take into account the different $(n,\,k)$ data used in this work. The air and c-Si layer are treated as infinite half spaces in the simulation. All layer thicknesses except the one of the perovskite are kept constant during the simulation. All values are in nm. (c) A layer stack with a pyramidally-textured rear side, which was used as reference in Section \ref{sec:lambertian}.} 
\label{fig:layerstacks}
\end{figure}

Figure \ref{fig:layerstacks}(a) shows the layer stack used for the simulation. Using the software package \texttt{JCMsuite}, which is based on the \emph{finite element method} (FEM), a three-dimensional mesh is created. The maximum side-length of the prismoidal or tetrahedral elements constituting the mesh is set to $\lambda/n(\lambda)$, where $\lambda$ is the incident wavelength and $n(\lambda)$ is the refractive index of the material. In this way one grid was generated for every wavelength interval of 100~nm width. More details can be found in the Section \ref{sec:convergence}, where we discuss a convergence study to justify the values used for the final simulations.

The layer stacks with planar perovskite top cells, which we discuss in Section \ref{sec:lambertian}, were simulated with \texttt{GenPro4}. This program is developed at Delft University of Technology and can combine wave optics for coherent thin layers and ray optics for thick incoherent layers \cite{santbergen:2017}. For planar devices with infinitely thick silicon absorbers results produces with \texttt{JCMsuite} and \texttt{GenPro4} match well\cite{chen:2018}.

The complex refractive index spectra ($n$, $k$) used for the simulations were determined as follows: perovskite data were retrieved using ellipsometry and transmittance/reflectance spectrophotometry \cite{guerra:2017}. For \ce{NiO_x} \cite{you:2016} and the sputtered ITO layers, ellipsometry and the program \texttt{RIGVM} was used \cite{pflug:2004}. The data for PCBM was extracted from reflectance/transmittance measurements with the method described in Refs.\ \citenum{albrecht:2014phd, djurisic:2000}. The \ce{SnO2} layers were deposited using plasma-enhanced atomic layer deposition and characterized with ellipsometry \cite{chistiakova:2018}. The data for the RF-PECVD hydrogenated amorphous silicon (a-Si:H) layers\cite{mazzarella:2017} were extracted using \texttt{SCOUT} \cite{scout}. For LiF \cite{li:1976} and Spiro-OMeTAD\cite{filipic:2015} we used data from literature. 

\subsection{Hexagonal Sinusoidal Nanotextures}
\label{sec:sinusoidal}

The \emph{hexagonal sinusoidal nanotextures} can be described with the equation
\begin{equation} \label{eq:hex}
\mathcal{A}(x,y;\phi,h) = \frac{h}{H(\phi)} \cos \left[\frac{1}{2}\left(x+\sqrt{3}y\right)\right]\cos\left[\frac{1}{2}\left(x-\sqrt{3}y\right)\right]\cos(x+\phi),
\end{equation}
where the \emph{structure phase} $\phi$ allows us to shape the hexagonal texture. Setting $\phi = 0$ leads to what we call a \emph{positive cosine} (``$+$cos'') structure whereas $\phi = \pi$ shapes the \emph{negative cosine} (``$-$cos'') structure, which is illustrated in Fig.\ \ref{fig:layerstacks}(a). The \emph{desired} peak-to-valley height $h$ is set as input value; $H(\phi)$ is the peak-to-valley height of the non-normalized texture\cite{jaeger:2016opex}
\begin{equation}
  H(\phi) = \frac{3\sqrt{3}}{4}\sin\left(\frac{\phi+\pi}{3}\right).
\end{equation}
For the structure phases used in this work we have $H(0) = H(\pi) = \frac{9}{8}$.

The function is scaled to our desired period $P$, which is the side length of the rhombus-shaped unit cell, with the substitutions
\begin{equation} \label{eq:sub}
x \rightarrow \frac{2\pi}{\sqrt{3}P}x
\quad\mathrm{and}\quad 
y \rightarrow \frac{2\pi}{\sqrt{3}P}y .
\end{equation}
The aspect ratio $a$ is defined as $a=h/P$; small or big values of $a$ lead to rather flat or steep textures, respectively. Note that we only considered \emph{negative cosine} structures in Ref.\ \citenum{chen:2018}.

\section{RESULTS}
\label{sec:results}

\subsection{Validity of Lambertian Approximation}
\label{sec:lambertian}

In the FEM simulations presented in Ref.\ \citenum{chen:2018} we treated the Si layer as an infinitely thick layer represented by a \emph{perfectly matched layer} (PML). For estimating the absorption in a 160~\textmu m-thick Si layer with perfect light trapping, we multiplied the absorption for the infinite Si layer $A_\text{Si}^\infty$ (equivalent to the transmission into this layer) with a factor arising from the \emph{Tiedje-Yablonovitch limit} \cite{tiedje:1984},
\begin{equation}
\label{eq:tiedje}
A_\text{Si}(\lambda,L) = A_\text{Si}^\infty(\lambda)\frac{\alpha(\lambda)}{\alpha(\lambda)+\left(4 \left[n_\text{Si}(\lambda)\right]^2L\right)^{-1}}.
\end{equation}
$L$ and $n_\text{Si}$ are the layer thickness and the real part of the refractive index of silicon, respectively. Further, we assumed the absorption of all other layers to be unaffected; the difference between the absorption in the infinite Si layer and the Tiedje-Yablonovitch limit was added to the reflectivity $R$,
\begin{equation}
	R(\lambda,L) = R^\infty(\lambda) + A_\text{Si}^\infty(\lambda) - A_\text{Si}(\lambda,L).
\end{equation}

In order to validate the accuracy of this approach, we compared the absorption spectra of the following two cases to each other: first, a flat perovskite-silicon solar cells with an infinitely thick Si layer [see Fig.\ \ref{fig:layerstacks}(b)], where the Si absorption was corrected according to Eq.\ (\ref{eq:tiedje}), here referred to as \emph{Tiedje}; secondly, a cell with a flat perovskite front side and a 160~\textmu m Si wafer with a pyramidal rear side, as in \ref{fig:layerstacks}(c) [see also Ref.\ \citenum{jaeger:2017eupvsec}]. This design we refer to as \emph{pyramidal}.

\begin{figure}
\centering
\includegraphics{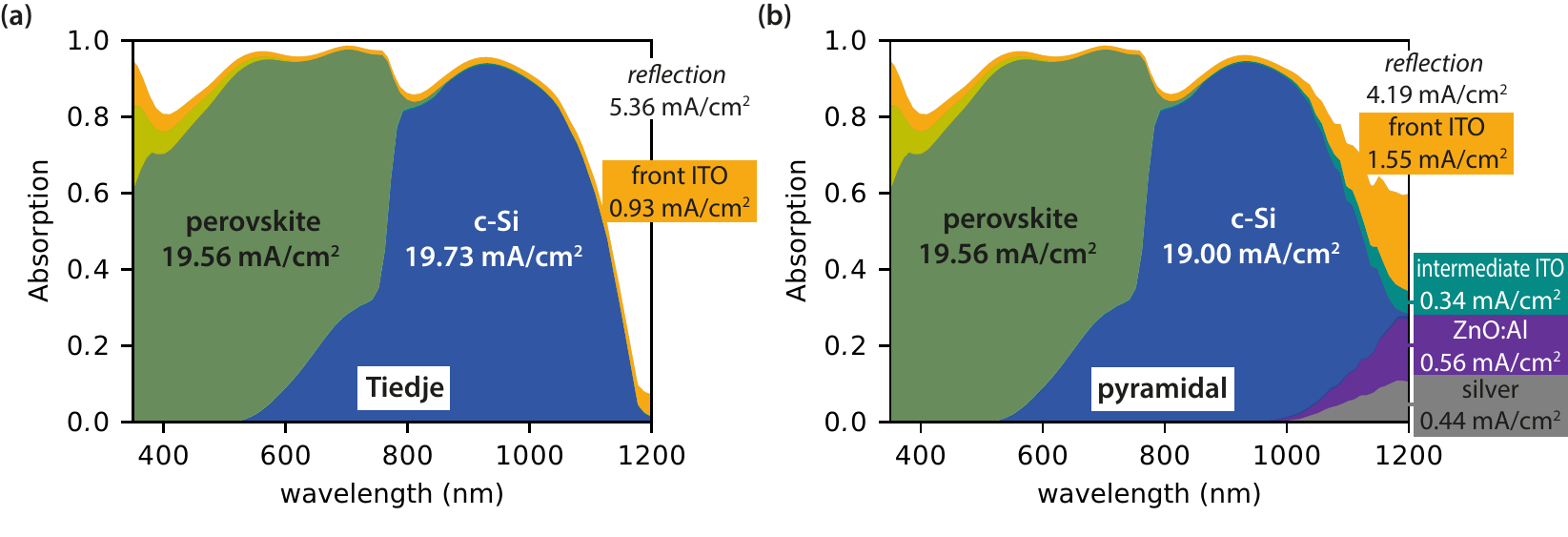}
\caption{Absorption profiles of (a) a planar solar cell with an infinitely thick silicon absorber [see Fig.\ \ref{fig:layerstacks}(b)], where the Si absorption is corrected for 160 \textmu m layer thickness with optimal light trapping according to the Tiedje-Yablonovitch limit as in Eq.\ (\ref{eq:tiedje}); and (b) a solar cell with a planar front side and a pyramidally textured silicon back side, as illustrated in Fig.\ \ref{fig:layerstacks}(c).} 
\label{fig:abs_profiles}
\end{figure}

Figure \ref{fig:abs_profiles} shows the absorption spectra for these two cases. Up to around 1000~nm wavelength, hardly any differences are visible because no light reaches the back of the Si wafer. For longer wavelength the Si absorption is higher for the Tiedje case (19.73~mA/cm$^2$ vs. 19.00~mA/cm$^2$), which is not surprising because this is the upper limit for light trapping. On the other hand, the absorption in the front ITO layer is underestimated in the Tiedje case (0.93~mA/cm$^2$ vs.\ 1.55~mA/cm$^2$).

In the Tiedje case, only the light which passes the ITO layer on its way from air into silicon contributes to the absorption in ITO. In conrast, in the pyramidal case also the light reflected back from the silicon rear side contributes to the absorption in ITO. In addition, the pyramids on the rear scatter the light leading to a prolonged light path and hence increased absorption in all the layers. In the pyramidal case about 0.34~mA/cm$^2$ are absorbed in the intermediate ITO layer, mainly arising from the long wavelengths. In the Tiedje case, around 0.10~mA/cm$^2$ are absorbed in that layer, the corresponding absorption area is hardly visible in Fig.\ \ref{fig:abs_profiles}(b). In the pyramidal case, also the absorption of the layers at the rear side of the silicon bottom cell is taken into account, which adds up to about 1~mA/cm$^2$.

In summary we can say that our approximation based on the Tiedje-Yablonovitch limit is well suited to estimating the Si absorption that would be present in a realistic solar cell stack with a pyramidal back reflector. Although this limit slightly overestimates the absorption compared to the more realistic case, the difference is small enough that the estimate remains useful. This validates the use of this estimate in Ref.\ \citenum{chen:2018}.

On the other hand, this estimate should not be used to estimate how losses are distributed among parasitic absorption and reflection. Since parasitic absorption at the rear side is neglected and light reflected from the rear side does not contribute to parasitic losses in the top cell (note that this light cannot contribute to absorption in the perovskite layer due to that layer being non-absorbing at these wavelengths).

\subsection{Convergence Study}
\label{sec:convergence}
In order to determine the accuracy of the FEM results a convergence study was performed. For regions without strong singularities the error of FEM simulations is expected to be proportional to $({h}/{\lambda})^{p}$.\cite{burger:2013derivative}. Therefore, for a sufficiently small mesh size $h$, the error can be decreased exponentially by increasing the order $p$ of the polynomials approximating the solution of Maxwell's equations in each element. Increasing the polynomial order for every element is typically superfluous as the local error is not equal everywhere in the computational domain. Therefore a strategy whereby the local error is first estimated for each element and then the required polynomial order $p$ is set locally for each element should maximize accuracy while minimizing computational cost. 

\begin{figure}
\centering
\includegraphics[scale=0.75]{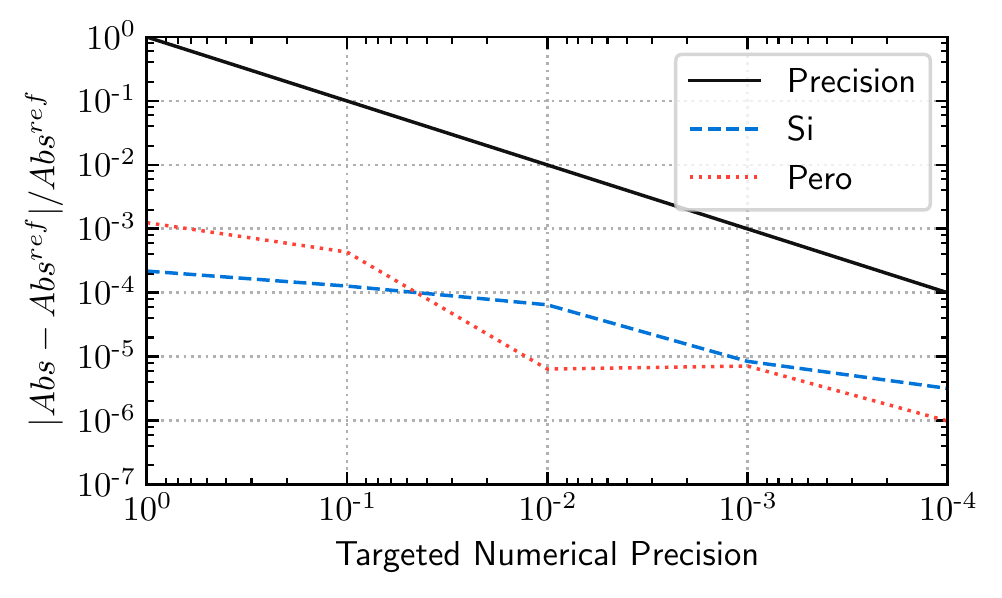}
\caption{The absolute relative error in the absorption in perovskite and Si layers as a function of the targeted numerical precision. Perovskite results are shown at 600~nm wavelength, while for Si 900~nm wavelength was used. As reference, a polynomial order $p=6$ for every element was used. Results were obtained for a nanotexture with 500~nm period and 250~nm height.} 
\label{fig:convergence}
\end{figure}

In order to determine the required order $p$, a targeted numerical precision should be given. Figure \ref{fig:convergence} shows the error in the absorption in both the Si and perovskite layers as a function of the given targeted numerical precision. All solutions are compared to a reference solution, where the polynomial order was $p=6$ for all elements. The absorption in the layers shows a much smaller error than the given numerical precision. This is because the numerical precision corresponds to the maximum error in the electric field over the whole domain. In comparison the absorption is calculated by integrating the field over the whole domain, which tends to reduce the effect of local errors. Nevertheless it is evident that as the targeted numerical precision is decreased, the error in absorption in the layers also decreases.

Hence, the numerical precision of 10$^{-3}$, which we used in Ref.\ \citenum{chen:2018}, leads to a sufficient level of accuracy for solar cell simulations.

\subsection{Comparing Different Sinusoidal Nanotextures}
\label{sec:nanotextures}

\begin{figure}
\centering
\includegraphics[scale=0.75]{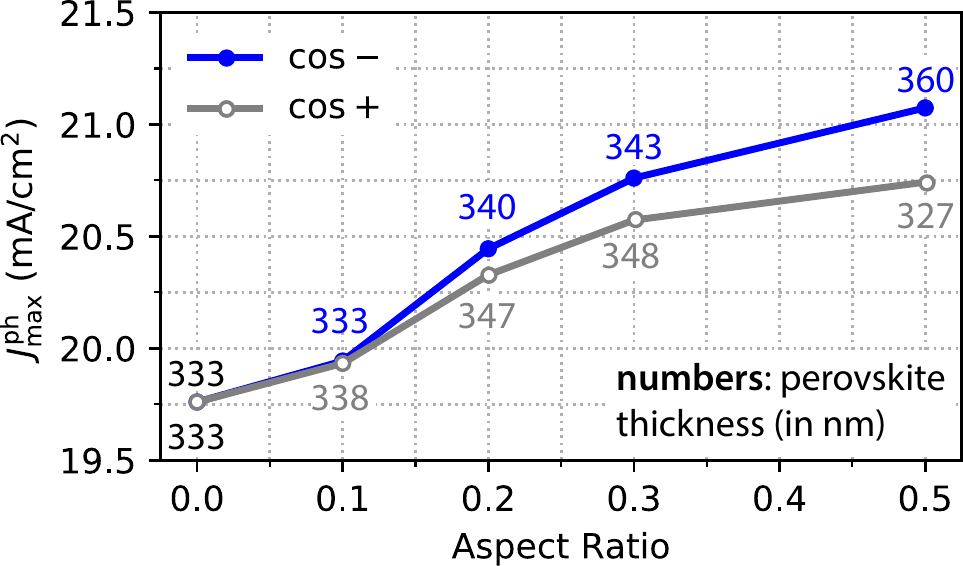}
\caption{The maximum achievable matched photocurrent density for the nanotextured tandem perovskite-Si solar cell for a positive and negative cosine texture as a function of the nanotexture aspect ratio. The nanotexture had 500~nm period. The thickness of the perovskite layer required for current matching is given as well.} 
\label{fig:textures}
\end{figure}

Figure \ref{fig:textures} shows the maximum achievable matched photocurrent density for the nanotextured perovskite-Si tandem solar cell as a function of the texture aspect ratio $a$. Shown are results for the positive and negative cosine textures, which we introduced in Section \ref{sec:sinusoidal}. Both textures present an increase in the matched photocurrent density upon increasing the aspect ratio. The matched current density increases due to the texture increasing the amount of light coupled into the device. The thickness of the perovskite layer needed for current matching in the case of the negative cosine texture increases monotonically with the aspect ratio. This suggests that the negative cosine texture transmits more light into the Si bottom cell, which is then balanced by increasing the perovskite thickness until the currents are matched.

In contrast the thickness of perovskite layer in the positive cosine structure initially increases more rapidly than in the case of the negative cosine before decreasing for aspect ratios up to 0.5. This suggests that positive cosine texture focuses more light inside the perovskite layer, increasing absorption and thereby necessitating a lower thickness for current matching. However this effect leads to an overall lower increase in the matched current compared to the negative cosine texture.

Therefore the negative cosine structure, utilized in Ref.\ \citenum{chen:2018}, is preferred for light management in perovskite Si tandem solar cells.

\section{CONCLUSIONS AND OUTLOOK}
We explained three details of the work presented in Ref.\ \citenum{chen:2018} in more depth. (1) We checked the validity of using the Tiedje-Yablonovitch limit for light trapping in the silicon subcell and saw that this estimate works well for the Si absorption but strongly underestimates the reflection in the long wavelength regime, because parasitic absorption in the other layers is strongly underestimated. (2) A convergence study for the mesh dimensions and finite element polynomial degree showed that the chosen settings lead to highly accurate results. (3) Comparing different configurations of the hexagonal nanotextures showed a similar trend as in earlier work (Ref.\ \citenum{jaeger:2016opex}).

In order to improve the simulation quality of solar cells with periodically nanotextured  structures and thick Si wafers, the coherent FEM simulations need to be coupled to approaches, which can treat thick layers incoherently. Such approaches are for example \texttt{OPTOS}\cite{hoehn:2018} or \texttt{GenPro4}.\cite{santbergen:2017}

\acknowledgments     
 
The numerical results were obtained at the \emph{Berlin Joint Lab for Optical Simulations for Energy Research} (BerOSE) of Helmholtz-Zentrum Berlin f\"{u}r Materialien und Energie, Zuse Institute Berlin and Freie Universit\"{a}t Berlin. 

K.J., D.E., G. K., and C.B. acknowledge the German Federal Ministry of Education and Research (BMBF) for funding the research activities of the Nano-SIPPE group within the program NanoMatFutur (grant no. 03X5520). 
P.M. is funded by the Helmholtz Innovation Lab HySPRINT, which is financially supported by the Helmholtz Association.
S.B. acknowledges support by Einstein Foundation Berlin through ECMath within subproject~OT9.
S.A. acknowledges the BMBF within the project “Materialforschung f\"{u}r die Energiewende” for funding of his Young Investigator Group (grant no. 03SF0540);  together with P.T. he acknowledges the German Federal Ministry for Economic Affairs and Energy (BMWi) for funding of the “PersiST” project (grant no. 0324037C).

We thank Johannes Sutter for conducting the AFM measurements and Florian Ruske for supporting us with the determination of optical properties.  



\end{document}